\documentclass[11pt,a4paper]{article}
\usepackage{amsfonts}
\newcommand{\vsp}{$\vphantom{\Big |}$}
\newcommand{\RR}{\mathbb{R}}
\newcommand{\CC}{\mathbb{C}}
\newcommand{\ZZ}{\mathbb{Z}}

\newcommand{\sco}[1]{\,[\![\,{#1}\,]\!]\,}
\newcommand{\D}{I\kern-3.5pt D}
\newcommand{\F}{I\kern-3.5pt F}
\newcommand{\Dsl}{\D\kern-7pt/\kern1pt}
\newcommand{\hf}{{\textstyle\frac{1}{2}}}
\newcommand{\oth}{{\textstyle\frac{1}{3}}}
\newcommand{\tth}{{\textstyle\frac{2}{3}}}
\newcommand{\tfrac}[2]{{\textstyle\frac{#1}{#2}}}
\newcommand{\bw}{{\textstyle\bigwedge}}

\newcommand{\arr}[1]{\smash{\mathop{\longrightarrow}\limits^{#1}}}
\newcommand{\CKM}{U_M}
\def\one{{\mathchoice {\rm 1\mskip-4mu l} {\rm 1\mskip-4mu l}
        {\rm 1\mskip-4.5mu l} {\rm 1\mskip-5mu l}}}
\begin{document}


\begin{center}
\LARGE\bf Towards a Unified Theory \\
          of Gauge and Yukawa Interactions              
\end{center}
\vspace{3mm}
\begin{center}\large 
       G.\ Roepstorff and Ch.\ Vehns\\
       Institute for Theoretical Physics\\
       RWTH Aachen\\
       D-52062 Aachen, Germany\\
       e-mail: roep@physik.rwth-aachen.de
\end{center}
\vspace{5mm}\par\noindent
{\bf Abstract}.
It is suggested to combine gauge and Yukawa interactions into one expression
$\overline{\Psi}\Dsl\Psi$ where $\Dsl$ is the generalized Dirac operator 
associated with a superconnection $\D=D+L$, $L$ being linked to the Higgs 
field (one doublet for simplicity).
\fontdimen16\textfont2=3.0pt
\fontdimen17\textfont2=3.0pt
We advocate a version of the Minimal Standard Model where the Higgs 
field gives masses to the neutrinos and a CKM matrix to the leptons.
Apart from a parameter $\mu\approx 80\,$GeV setting the mass scale, the
(dimensionless) free parameters of three fermion generations
are assembled in one operator $h$, invariant under gauge transformations. 
As we are free to choose $h$, the predictive power is rather limited. 
Still, the fine structure constant and the weak couplings remain unaffected:
$\alpha^{-1}=128\pi/3$, $g=1/2$, $g'/g=\sqrt{3/5}$. There are three 
relations that fix the masses of the $Z^0$, the $W^\pm$, and the Higgs, 
given the masses of all fermions. The present data are consistent with
$m_H\approx 160\,$GeV. Without these data, $m_H\ge\sqrt{2}\,m_W$ on general
grounds.

\section{Introduction}

It is fair to say that, at present, the Standard Model belongs to the
category of most thoroughly tested and best confirmed physical theories [1]. 
But the secret is that no one truly understands it. At least not in the way we
understand QED or QCD. It appears that the vast number of
free parameters, speculations about the number of Higgs doublets, and
the ad-hoc definition of the Higgs potential defy easy analysis.
With each new generation of fundamental fermions the unknowns have
multiplied. The neutrinos may or may not have masses. It remains undecided
whether nature provides another CKM matrix for the leptons. 
In the long run, the struggle for a better understanding will perhaps 
be resolved by resorting to string theory (M-theory or other oracles). 
In the meantime we might be content with modest explanations using 
constructions
in ordinary (commutative) differential geometry. One concept, which
convincingly illuminates the role of the Higgs field, comes under the heading
{\em superconnection}. Implied is the concept of a generalized Dirac operator
which unites gauge and Yukawa couplings in one term. 

In [2] we showed how the Higgs field fits into the 
framework of superconnections on some superbundle with structure group
$U(n)$. We then proposed to take $n=2$ to construct a model
of gauge fields and two Higgs doublets. In [3] we added leptons to the model
to see how gauge and Yukawa interactions can be combined by passing from
a superconnection to the associated Dirac operator. Constructing such a model 
was a tentative step towards an understanding of the structure of more 
realistic theories. 
Of course, without quarks the leptonic model was not free of anomalies and thus called 
for an extension incorporating essential features of the Standard Model. 
A short account of such an attempt appeared in [4] where the gauge group $G$
was assumed to be a subgroup of $SU(5)$. Parallel to this work we opened 
a discussion in [5] and [6] on the mathematical background which should serve 
as a reference when we now resume the analysis of the Standard Model
begun in [4]. Moreover, since we are dealing here with a chiral 
model, the question of consistency (absence of local nonabelian anomalies) 
arises. This problem has been dealt with and settled in [7].

The present exposition of the subject aims to provide motivation and
practical tools rather than mathematical abstraction. Its goal is
to arrive at predictions with minimal technical machinery.

\section{The Gauge Group of the Standard Model}

The Standard Model, extending of the earlier Weinberg-Salam model of
electroweak interactions, is a gauge theory based on the Lie algebra
\begin{equation}
 \mbox{Lie\,}G \cong{\bf su}(3)\oplus{\bf su}(2)\oplus{\bf u}(1)\ . \label{Lie}
\end{equation}
Though the gauge group $G$ is well defined locally, its global
structure remains obscure unless we add further assumptions.
Obviously, no restriction on the spectrum of the hypercharge $Y$ is 
to be expected on the basis of ${\bf u}(1)=i\RR$ alone. Nor would one be able
to argue that the conditions $Q+Y\in\ZZ$ and $3Y\in\ZZ$ are satisfied where
$Q$ denotes the electric charge. Still, in the vicinity of the unit, 
a possible difference between various choices of $G$ would not be felt at all, 
but globally it would: $G$ will dictate the subset of allowed representations
of $\mbox{Lie\,}G$ and, therefore, the structure of particle multiplets
admitted by the setup.

The present approach borrows from the idea that any grand unified
theory, perhaps any theory beyond the Standard Model, ought to encorporate
the gauge group $SU(5)$ in one way or another. As for the minimal version of
the Standard Model, we require that $G$ be a subgroup of $SU(5)$ consistent
with (\ref{Lie}), i.e., we define
\begin{equation}
    G =\{(u,v)\in U(3)\times U(2)\ |\ \det u\cdot\det v=1\}
\end{equation}
and let the embedding $G\to SU(5)$ be given by
$$
                  (u,v) \mapsto\pmatrix{u &0\cr 0&v\cr}\ .
$$
It is easy to see that the Lie algebra $\mbox{Lie\,}G$ has indeed the required
structure given by (\ref{Lie}).

The relation of the group $G$ to the color group $SU(3)$ and the electroweak
group $U(2)$ is expressed by the following exact sequence
\begin{equation}
     1\arr{}SU(3)\arr{j}G\arr{s}U(2)\arr{}1     \label{exs}
\end{equation}
where $j(u)=(u,1)$ and $s(u,v)=v$. Inspite of the relationship (\ref{exs}), 
the group $G$ cannot be identified with the direct product $SU(3)\times U(2)$. 
It is still correct to say that the color group $SU(3)$ of quantum 
chromodynamics is embedded in $G$ as a subgroup. But the gauge group 
$U(2)$ of the Salam-Weinberg Theory is recovered here only as the quotient 
$G/SU(3)$. This fact deserves careful attention: it explains why the 
hypercharge $Y$ assumes fractional values. More specifically, the quotient 
structure accounts for the existence of values that are multiples of 1/3.

The theory of leptons, gauge and Higgs particles is built around the 
assumption that the hypercharge $Y$
is the generator of $U(1)$, subgroup and center of $U(2)$. As these groups 
constitute proper symmetries (before spontaneous symmetry breaking takes 
place), the hypercharge is integer-valued for all leptonic states. When quarks
are added, the picture changes. With quarks, the group $U(1)$ fails to be
a subgroup of $G$ and hence cannot be regarded a symmetry though there is
a related group $\tilde{U}(1)$ which can. To see more clearly
the emergence of a fractional spectrum we consider the exact sequence
\begin{equation}
  \label{eq:der}
  1\arr{}\ZZ_3\arr{j}\tilde{U}(1)\arr{s}U(1)\arr{}1
\end{equation}
obtained from the groups in (\ref{exs}) by restricting to the centers. In more detail:
\begin{itemize}
\item The center $\tilde{U}(1)$ of $G$ consists of elements
 $$   
      \mbox{diag}(e^{i\beta},e^{i\beta},e^{i\beta},e^{i\alpha},e^{i\alpha})
      \in SU(5)
 $$
satisfying $(e^{i\beta})^3(e^{i\alpha})^2=1$. It may conveniently be looked 
upon as a one-dimensional closed subgroup of the two-torus: 
$$
  \tilde{U}(1) =\{(e^{i\beta},e^{i\alpha})\ |\ 3\beta+2\alpha=
                                               0 \bmod 2\pi\}. 
$$
\item The cyclic group $\ZZ_3$ of order three is formed by the complex
solutions of $z^3=1$. The injection $j$ takes $z$ into 
$(z,1)\in\tilde{U}(1)$.
\item The surjection $s$ maps $(e^{i\beta},e^{i\alpha})$ to 
$e^{i\alpha}\in U(1)$ where $U(1)$ relates to the hypercharge.
\end{itemize}
Viewed geometrically, the group $\tilde{U}(1)$ describes a closed curve
on the 2-torus. Suppose we unwind the torus to obtain its covering plane
with real coordinates $\alpha$ and $\beta$. Then the curve appears as a
straight line with slope parameter $-2/3$:
$$
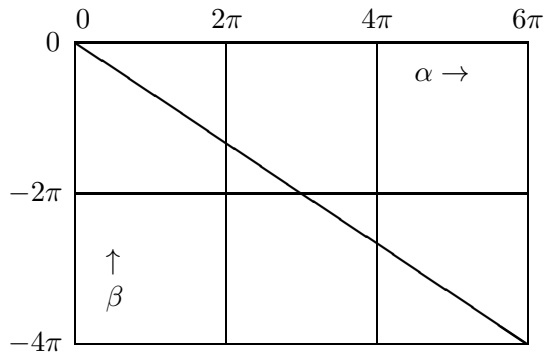
\unitlength1cm
\begin{picture}(6,5)
   \put(0,0){\framebox(6,4){}}
   \put(2,0){\line(0,1){4}}
   \put(4,0){\line(0,1){4}}
   \put(0,2){\line(1,0){6}}
   \thicklines
   \put(0,4){\line(3,-2){6}}
   \put(0,4.2){0}
   \put(1.8,4.2){$2\pi$}
   \put(3.8,4.2){$4\pi$}
   \put(5.8,4.2){$6\pi$}
   \put(-0.4,3.9){0}
   \put(-0.9,1.9){$-2\pi$}
   \put(-0.9,-0.1){$-4\pi$}
   \put(4.5,3.5){$\alpha\rightarrow$}
   \put(0.4,1.0){$\uparrow$}
   \put(0.4,0.5){$\beta$}
\end{picture}\vspace{8mm}
$$
The end points of the line have to be identified to
form a closed curve on the torus. As we run once through this curve,
the angle $\alpha$ assumes all values from 0 to $6\pi$. Phrased more formally,
$\tilde{U}(1)$ is a threefold cover of the group $U(1)=\{e^{i\alpha}\}$.
When it comes to particle multiplets, we must focus on the symmetry group 
$\tilde{U}(1)$ since its generator has an integer-valued
spectrum. In unitary irreducible representations of the gauge group $G$,
the hypercharge $Y$ assumes a constant value subject to the constraint
$$
          e^{i6\pi Y}=1\qquad\mbox{oder}\qquad 3Y\in\ZZ\,,
$$
as can be inferred from the behavior of the variable $\alpha$.
In other words, the group $\tilde{U}(1)$
is not connected in a direct manner with $Y$ but rather with $3Y$. Phrased
more formally, the covering map $s: \tilde{U}(1)\to U(1)$
admits a local inverse
\begin{equation}
  \label{eq:inv}
      s^{-1}(e^{i\alpha})=(e^{-i2\alpha/3},e^{i\alpha}).
\end{equation}
and hence, at least locally (for small $\alpha$), the group $U(1)$ is
represented by a phase factor $e^{-i\alpha Y}$ in any unitary irreducible 
representation of $G$, in such a way that $3Y$ becomes an integer.

In the leptonic sector, spontaneous symmetry breaking selects another 
one-parameter subgroup of $U(2)$ which remains unbroken and gives rise 
to the concept of electric charge. By convention, this subgroup is
\begin{equation}
  \label{ucha}
        U(1)_Q =\left\{\pmatrix{1&0\cr 0&e^{i\alpha}\cr}\right\}  
\end{equation}
Therefore, the charge $Q$ assumes integer values in this sector. Again,
with quarks the situation changes. The group $U(1)_Q$ being no longer
a symmetry is replaced by its threefold cover,
\begin{equation}
  \label{ch2}
  \tilde U(1)_Q=
  \left\{ (u,v)\ \Big|\ u=e^{i\beta}\one_3,\ 
        v=\pmatrix{1&0\cr 0&e^{i\alpha}\cr},\ 3\beta+\alpha=0\bmod 2\pi\right\}
\end{equation}
Locally, we may put $\beta=-\alpha/3$ and $\tilde U(1)_Q=e^{-i\alpha Q}$ such
that $3Q\in\ZZ$.

Summarizing, the hypercharge and the electric charge are represented on
$\CC^5$ by the following traceless matrices:
\begin{eqnarray*}
               Y &=& \mbox{diag}(\tth,\tth,\tth,-1,-1)\\
               Q &=& \mbox{diag}(\oth,\oth,\oth,0,-1)
\end{eqnarray*}
Though many states, especially those invariant under the color group $SU(3)$,
carry integer charges, $Q$ and $Y$ assume fractional values in general, 
still obeying $Q+Y\in\ZZ$.

\section{The $G$-Supermodule of Fermions}

By construction, there is a natural irreducible unitary action of the gauge 
group $G$ on the space
\begin{equation}
  \label{c3c2}
            \CC^5=\CC^3\oplus\CC^2
\end{equation}
with subspaces $\CC^3$ and $\CC^2$ carrying {\em fundamental 
representations\/} of the color group SU(3) and the weak-isospin group $SU(2)$
respectively. But, passage to the $\ZZ_2$-graded exterior algebra 
$$
    \bw\CC^5=\sum_{k=0}^5 \bw^k\CC^5=\bw^+\CC^5\oplus\bw^-\CC^5\,,\qquad\quad
    \bw^{\pm}\CC^5=\sum_{(-1)^k=\pm1}\bw^k\CC^5
$$
is very essential if we want to let $G$ act on a superspace. For a general
discussion and details concerning the exterior algebra as a superspace we 
refer to [5]. It is apparent that the induced unitary representation $\bw$
of $G$ on $\bw\CC^5$ is reducible. We take the view that $\bw\CC^5$ is
the basic $G$-module for fermions. Quarks and leptons of one generation will be
grouped according to the irreducible constituents of the representation $\bw$.
In addition, there is another space $\CC^3$, not a $G$-module, which describes
the flavor degrees of freedom. While we
refer to $\bw\CC^5$ as the {\em $G$-supermodule of fermions}, we call the
tensor product
$$
          \bw\CC^5\otimes\CC^3          
$$
the {\em inner space}, because it incorporates all inner degrees of freedom.

We now turn to the structure of the $G$-supermodule.
From (\ref{c3c2}) and the natural isomorphism (of vector spaces)
$$
     \bw(\CC^3\oplus\CC^2)\cong\bw\CC^3\otimes\bw\CC^2\,,
$$
where
$$\textstyle
  \bw\CC^3=\sum_{p=0}^3\bw^p\CC^3\,,\qquad \bw\CC^2=\sum_{q=0}^2\bw^q\CC^3\,,
$$
we obtain $\bw(u,v)=\bw u\otimes\bw v$  for $(u,v)\in G$ and hence
$$
 \textstyle \bw^k(u,v)=\sum_{p+q=k}\bw^pu\otimes\bw^q v,\qquad k=0,\ldots,5.
$$
We call
\begin{equation}
  \label{pari}
                 \kappa=(-1)^k=(-1)^{p+q}
\end{equation}
the {\em parity operator\/} in $\bw\CC^5$.

Within a generation of fermions, each multiplet (left- or 
right-handed) is associated with one of the 
following irreducible representations of $G$,
$$       
   \bw^{p,q}=\bw^p\otimes\bw^q\qquad p=0,1,2,3,\qquad q=0,1,2
$$ 
whose dimension is ${p\choose 3}{q\choose 2}$.
To find its hypercharge we use Eq.\ (\ref{eq:inv}),
$$ 
    e^{-i\alpha Y} = \bw^{p,q}\big(s^{-1}(e^{i\alpha})\big)
                   =\exp(-i2p\alpha/3+iq\alpha)\,,
$$
and thus obtain the fundamental relation
\begin{equation}
  \label{eq:pq}
              \textstyle Y =\frac{2}{3}p-q\ .
\end{equation}
We distinguish
$$
\begin{tabular}{ll}
      lepton fields &: $p=0$ or 3\\
      quark fields  &: $p=1$ or 2.
\end{tabular}
$$
The electric charge then satisfies the formula of Gell-Mann-Nishijima
$$
                 Q=I_3+\hf Y
$$
where $I_3$ denotes the third component of the weak isospin. Clearly,
$I_3=0$ if $q=0,2$ and $I_3=\pm\hf$ if $q=1$.

With each generation we associate a generalized Dirac field $\psi$ having 
$2^5=32$ elementary Weyl spinors as its components. Spinors that enter $\psi$ 
are characterized by three different ``parities'' owing to the $\ZZ_2$-gradings
of $\bw\CC^5$, $\bw\CC^3$, and $\bw\CC^2$. Their interpretation is as follows
(recall that $k=p+q)$): 
$$
\begin{tabular}{rlrl}
  $k = $even  &: right-handed \hspace{20mm} &$k   = $odd &: left-handed\\
  $p = $even  &: matter                     &$p   = $odd &: antimatter\\
  $q = $even  &: singlets                   &$q   = $odd &: doublets.
\end{tabular}
$$
Since charge conjugation acts on the inner space by complex conjugation, it
passes from $[p,q]$ to $[3-p,2-q]$. It thus interchanges left and right,
matter and antimatter, and reverses the signs of $Y$, $I_3$ and $Q$, but
takes singlets into singlets and doublets into doublets.

\section{Choosing a Basis in $\bw\CC^5$}

To describe the field $\psi$ in more conventional terms we need to 
construct a basis of eigenvectors in $\bw\CC^5$. Such a construction
starts from a basis in $\CC^5$. Let us emphazise: 
besides being a 5-dimensional complex
linear space, $\CC^5$ is endowed with a Hermitian structure and also comes 
with a distinguished orthonormal basis $(e_i)_{i=1}^5$ such that 
$(e_i)_j=\delta_{ij}$. We shall refer to it as the {\em standard basis\/}
in $\CC^5$.

An induced basis $e_I$ in $\bw\CC^5$ is then given by 
$$
    e_I =e_{i_1}\wedge\cdots\wedge e_{i_k}\in\bw^k\CC^5,\qquad 
    I=\{i_1,\ldots i_k\},\quad i_1<\cdots<i_k,\quad 0\le k\le 5
$$
where $I$ runs over all subsets of $\{1,2,3,4,5\}$ including the empty set 
$\emptyset$. We simply have to identify the numbers $k,p,q$ previously introduced.
Given any subset $I$,
\begin{itemize}
\item $k$ is the number of elements in $I$ taken from $\{1,2,3,4,5\}$, 
\item $p$ is the number of elements in $I$ taken from $\{1,2,3\}$,
\item $q$ is the number of elements in $I$ taken from $\{4,5\}$.
\end{itemize}
We shall also write $|I|$ in place of $k$, and
the complement of $I$ in $\{1,2,3,4,5\}$ is denoted $I^c$.
\par\noindent
The following table lists all 32 basis vectors and groups them according to 
their $p$ and $q$ values. For convenience, we write $I$ where we really 
mean $e_I$.\vspace{3mm}
\begin{equation}\label{ind}
\begin{tabular}{|c|rr|rr|}\hline
   $I$   & $q=0$           & $q=2$ & \multicolumn{2}{c|}{$q=1$} \\ \hline
   $p=0$ &\vsp $\emptyset$ &   45  &4 & 5\\  \hline
         &\vsp  23 & 2345  &  234  &  235\\
   $p=2$ &\vsp  13 & 1345  &  134  &  135\\
         &\vsp  12 & 1245  &  124  &  125\\ \hline \hline
         &\vsp   1 &  145  &   14  &   15\\
   $p=1$ &\vsp   2 &  245  &   24  &   25\\
         &\vsp   3 &  345  &   34  &   35\\ \hline
   $p=3$ &\vsp 123 & 12345 & 1234  & 1235\\ \hline
\end{tabular}
\end{equation}
\vspace{3mm}

The basis vectors of the first two columns have $I_3=0$ while those of the 
third and fourth column have $I_3=\hf$ and $I_3=-\hf$ respectively.
By construction, each basis vector $e_I$ is an eigenvector of $Y$ and $Q$.

The following two tables provide the hypercharges (left table) and
the electric charges (right table) associated with the basis vectors:
\vspace{3mm}
$$
\begin{tabular}{|c|rr|rr|}\hline
    $Y$  & $q=0$    & $q=2$ & \multicolumn{2}{c|}{$q=1$} \\ \hline
   $p=0$ &\vsp $0$  & -2    & -1     & -1  \\  \hline
         &\vsp 4/3  & -2/3  & 1/3    & 1/3 \\
   $p=2$ &\vsp 4/3  & -2/3  & 1/3    & 1/3 \\
         &\vsp 4/3  & -2/3  & 1/3    & 1/3 \\ \hline \hline
         &\vsp 2/3  & -4/3  & -1/3   & -1/3\\
   $p=1$ &\vsp 2/3  & -4/3  & -1/3   & -1/3\\
         &\vsp 2/3  & -4/3  & -1/3   & -1/3\\ \hline
   $p=3$ &\vsp 2    & 0     & 1      & 1   \\ \hline
\end{tabular}\hspace{3.5mm}
\begin{tabular}{|c|rr|rr|}\hline
   $Q$   & $q=0$    & $q=2$ & \multicolumn{2}{c|}{$q=1$} \\ \hline
   $p=0$ &\vsp $0$  & -1    & 0      & -1  \\  \hline
         &\vsp 2/3  & -1/3  & 2/3    & -1/3\\
   $p=2$ &\vsp 2/3  & -1/3  & 2/3    & -1/3\\
         &\vsp 2/3  & -1/3  & 2/3    & -1/3\\ \hline \hline
         &\vsp 1/3  & -2/3  & 1/3    & -2/3\\
   $p=1$ &\vsp 1/3  & -2/3  & 1/3    & -2/3\\
         &\vsp 1/3  & -2/3  & 1/3    & -2/3\\ \hline
   $p=3$ &\vsp 1    & 0     & 1      &   0 \\ \hline
\end{tabular}
$$
\vspace{3mm}

After this assignment of charges, the basis vectors $e_I$ can be put in a
1:1 correspondence with the 32 Weyl spinors of the first generation:
\begin{equation}
\begin{tabular}{|c|rr|rr|}\hline
 1.\ gener. & $q=0$          & $q=2$      & \multicolumn{2}{c|}{$q=1$}\\ \hline
 $p=0$ &\vsp $\nu_{eR}$ & $e_R$      & $\nu_{eL}$ & $e_L$        \\ \hline
       &\vsp $u_{1R}$   & $d_{1R}$   & $u_{1L}$   & $d_{1L}$     \\
 $p=2$ &\vsp $-u_{2R}$  & $-d_{2R}$  & $-u_{2L}$  & $-d_{2L}$    \\
       &\vsp $u_{3R}$   & $d_{3R}$   & $u_{3L}$   & $d_{3L}$  \\ \hline \hline
       &\vsp $d^c_{1L}$ & $u^c_{1L}$ & $d^c_{1R}$ & $-u^c_{1R}$  \\
 $p=1$ &\vsp $d^c_{2L}$ & $u^c_{2L}$ & $d^c_{2R}$ & $-u^c_{2R}$  \\
       &\vsp $d^c_{3L}$ & $u^c_{3L}$ & $d^c_{3R}$ & $-u^c_{3R}$  \\ \hline
 $p=3$ &\vsp $e^c_L$    & $\nu^c_{eL}$ & $e^c_R$  & $-\nu^c_{eR}$\\ \hline
\end{tabular}
\label{tabu}
\end{equation}
Two more tables of this kind exist for the second and third generation.
Some remarks are in order:
\begin{itemize}
\item The symbols (including their sign) stand for the components $\psi_I$
      of the field $\psi$ where the appropriate subset $I\subset
      \{1,2,3,4,5\}$ is displayed in the table (\ref{ind}). For instance,
      $$    \psi_{13}=-u_{2R}\qquad\mbox{etc.} $$
      By assumption, all fermions are massless to begin with.
      Therefore, each entry also represents a particle or antiparticle.
      The upper half of the table contains 
      the particles (``matter'') while the lower half contains the 
      antiparticles (``antimatter''). 
\item Quarks such as $u$(up) and $d$(down) come in three {\em colors\/}: 
      $i=1,2,3$. While quarks transform under the color group according to
      the representation $\bar{3}$, antiquarks transform according to the
      representation 3. Since both 3 and $\bar{3}$ are fundamental irreps
      of $SU(3)$, interchanging their role, as done here, has no physical 
      effect.
\item Together with each spinor the charged conjugate spinor (with upper index
      $\vphantom{d}^c$) also enters the table (\ref{tabu}) and so enters the
      field $\psi$. Since charge conjugation changes the chirality, 
      we have to make precise what the symbols in (\ref{tabu}) really mean. 
      With the $d$-quark as an example our convention is:
      $$ d^c_L :=(d^c)_L =(d_R)^c,\qquad d^c_R:=(d^c)_R=(d_L)^c\,. $$ 
\item Contrary to the traditional formulation of the Minimal Standard Model,
      there is room for a right-handed neutrino. Note the presence of the
      unconventional field $\nu_{eR}$ (together with its charge conjugate
      $\nu^c_{eL}$) in the table (\ref{tabu}) and the fact that it transform 
      trivially under the gauge group $G$.
      Thus, the right-handed neutrino does not couple to any gauge field
      whatsoever which makes it hard to detect it in experiments.
      It only couples to the Higgs field and so acquires a mass
      after symmetry breaking.
\item Algebraic reasoning has led us to include certain minus signs in the
      table (\ref{tabu}). One of the reasons is that we want the following
      condition to be satified:
      \begin{equation}
             \sigma_I \psi_I^c = \psi_{I^c}  \label{Ic}
      \end{equation}
      where $\sigma_I$ is the sign of the permutation taking $\{I,I^c\}$ to its
      normal order $\{1,2,3,4,5\}$. This in particular guarantees that, if
      $(\nu_{e},e)_L$ transforms as a $SU(2)$ doublet, so does the pair
      $$
                (e^c, -\nu_e^c)_R =(e_L, -\nu_{eL})^c
      $$
      after charge conjugation.
\end{itemize}

\section{The Concept of a Generalized Majorana Field}

The field $\psi$ is thought of as some generalized Dirac field having 
sufficiently many components $\psi_I$ to as to be able to describe all 
fundamental fermions of one generation. For its mathematical construction 
we need to introduce the spinor space $S$, basic to any Dirac field. 
Since its structure depends merely on the choice of spacetime (of even 
dimension in any case), we call $S$ the {\em outer space\/}.
In the language of [4], $S$ is a {\em Clifford supermodule\/}, i.e., 
a linear space on which the $\gamma$ matrices act, carrying 
a $\ZZ_2$-grading 
\begin{equation}
                     S=S^+\oplus S^-
\end{equation}
given by the chirality, the eigenvalues $\pm 1$ of $\gamma_5$. 
Field components taking values in $S^+$ ($S^-$) are said to be right-handed 
(left-handed).  

In addition, we have assumed that there is another superspace, $\bw\CC^5$,
graded by the parity $\kappa$ of exterior powers. This space, specific to
the Standard Model, is the same for all fermion generations.
Since we wish to relate the chirality in $S$ to the parity 
in $\bw\CC^5$, the field $\psi$ is required to take values in the even 
part of the tensor product $E=\bw\CC^5\otimes S$ which is
\begin{equation}\textstyle
               E^+=(\bw\CC^5\otimes S)^+ =
              (\bw^+\CC^5\otimes S^+)\,\oplus\,(\bw^-\CC^5\otimes S^-)\ .
\end{equation}
In order to write $\psi$ in terms of its components $\psi_I$ we use the basis 
$e_I$ for
the inner space as constructed in the previous section. With respect to the 
tensor product $\bw\CC^5\otimes S$, we decompose the field $\psi$ as
\begin{equation}
               \psi(x) =\sum_I e_I\otimes\psi_I(x) \ \in\  E^+\,,\qquad
               \psi_I(x)\in S\,.
\end{equation}
The condition $\psi\in E^+$ translates into
\begin{equation}
                   \gamma_5\psi_I =(-1)^{|I|}\psi_I\ .
\end{equation}
Hence, $\psi_I$ is right(left)-handed depending on whether $|I|$ is even(odd).

Two operations of similar nature, one in $\bw\CC^5$ and one in $S$, will play 
an important role:
\begin{itemize}
\item The Hodge operator $*:\bw\CC^5\to\bw\CC^5$ is antilinear and acts on
      the basis as
      \begin{equation}
             *e_I=\sigma_I e_{I^c}\ . \label{eI}
      \end{equation}
      Recall that $\sigma_I$ is the sign of the permutation taking $\{I,I^c\}$
      to its
      normal order $\{1,2,3,4,5\}$. Since $|I|+|I^c|=$odd,
      the Hodge operator is parity changing: 
      \begin{equation}
        \label{skap}
                 *\kappa*=-\kappa\,.        
      \end{equation}
      Note that $\sigma_{I^c}=\sigma_I$ (valid in odd dimensions) making
      the Hodge $*$ an involutive operator: $*^2=\one$
\item The charge conjugation $S\to S$, $s\mapsto s^c$, is antilinear, 
      involutive and reverses the chirality: $S^\pm\to S^\mp$.
\end{itemize}
The identification 
$$
       \mbox{parity in $\bw\CC^5$}\ =\ \mbox{chirality in $S$}
$$  
suggests to couple both operations, resulting in a single 
antilinear operator $*:E^+\to E^+$ such that
\begin{equation}
  \label{op}
        \psi(x)=\sum_Ie_I\otimes\psi_I(x) \qquad\Rightarrow\qquad
         *\psi(x)=\sum_I *e_I\otimes\psi^c_I(x)\ .     
\end{equation}
Though the $*$ operator now changes matter into antimatter and vice versa,
it should not be confused with ``charge conjugation'' in the traditional sense,
$$
                   \psi^c(x)=\sum_I e_I\otimes\psi^c_I(x)\,,
$$
which is not a symmetry. Using (\ref{Ic}) and (\ref{eI}) we find 
\begin{eqnarray*}
  *\psi(x) &=& \sum_I\sigma_I e_{I^c}\otimes\psi^c_I(x)
             = \sum_I e_{I^c}\otimes\sigma_I\psi^c_I(x)\\
           &=& \sum_I e_{I^c}\otimes\psi_{I^c}(x) 
             = \sum_Ie_I\otimes\psi_I(x)\ =\ \psi(x)
\end{eqnarray*}
A generalized Dirac field $\psi$ based on the $G$-module $\bw\CC^5$ is said to
be selfdual\footnote{The definition of selfduality works for spaces
$\bw\CC^n$ where $n$ is odd. If $n=1$, the concept reduces to that of
an ordinary Majorana field.} or a {\em generalized Majorana field\/} if it 
satisfies the relation $*\psi=\psi$.

Experimentally, three generations of fundamental fermions have been found:
from the decays of the $Z^0$ boson one infers that there are exactly three
generations (i.e., three is the number of neutrinos with masses below 45 GeV).
It is a trivial matter to combine the Dirac fields $\psi_f$ ($f=1,2,3$)
of three generations to a single {\em master field\/}:
\begin{equation}
  \label{master}
  \Psi =\psi_1\otimes\epsilon_1+\psi_2\otimes\epsilon_2
       +\psi_3\otimes\epsilon_3
\end{equation}
where $(\epsilon_i)_{i=1}^3$ denotes the standard basis in $\CC^3$, the 
flavor space. For con\-sisten\-cy, the $*$ operator must also act on $\CC^3$ in
an antilinear manner. We let it coincide with
complex conjugation. Hence, if $A$ is some matrix in $\mbox{End}\,\CC^3$,
then $*A*$ stands for the complex conjugate matrix. The fundamental relation 
$*\psi_f=\psi_f$, valid in each generation, is now equivalent to stating that
$*\Psi=\Psi$.

\section{Analysis of Operators}

The complex space $\CC^5$ we started from not only provides a natural basis
$(e_i)_{i=1}^5$ but is also equipped with a Hermitian structure given 
by the standard scalar product. 

Supposing $V$ is any Hermitian vector space, one associates a multiplication
operator $\epsilon(v)$ to any $v\in V$, acting on the exterior algebra,
$$
           \epsilon(v)a=v\wedge a\qquad(a\in\bw V),
$$
and lets $\iota(v)$ be its adjoint with respect to the induced Hermitian
structure in $\bw V$. In the language of Fock spaces, these operators are said
to be creation and annihilation operators respectively. It is evident that
they are of odd type, i.e., they change the parity of elements in $\bw V$.
We state this property as
$$
              \epsilon(v),\iota(v)\in\mbox{End}^-\bw V\ .
$$
With $V=\CC^5$ we put $b_i=\epsilon(e_i)$ and $b^*_i=\iota(e_i)$ so as to
obtain the usual anticommutation relations:
$$
     \{b_i,b^*_k\}=\delta_{ik},\quad \{b_i,b_k\}=0,\quad \{b^*_i,b^*_k\}=0.
$$
The key observation is that lifting matrices
$a=(a_{ik})\in\mbox{End}\,\CC^5$ to observables now admits an explicit 
description:
\begin{equation}\label{theta}
          \theta(a)=\sum_{ik}a_{ik}b_ib^*_k\ \in \mbox{End}^+\bw\CC^5. 
\end{equation}
To put it more formally, $\theta(a)$ is a {\em derivation\/} of the algebra 
$\bw\CC^5$. As for us, $\theta(a)$ is simply a linear operator of even type.
Being a $G$-module, $\bw\CC^5$ carries a representation $\theta$ of Lie$\,G$.
Note that the map $\theta$ constructed above extends the representation of
the Lie algebra.
 
Some of the previously introduced observables receive a 
new description:
\begin{eqnarray}
    Q &=&\oth(b_1b_1^*+b_2b_2^*+b_3b_3^*)-b_5b_5^*\nonumber\\
    Y &=&\tth(b_1b_1^*+b_2b_2^*+b_3b_3^*)-b_4b_4^*-b_5b_5^*\nonumber\\
   I_3&=&\hf(b_{4}b_4^*-b_{5}b_5^*)  \label{charge}\\
    p &=&b_1b_1^*+b_2b_2^*+b_3b_3^*\nonumber\\
    q &=&b_{4}b_4^*+b_{5}b_5^*\ .\nonumber
\end{eqnarray}
The significance of the operators $p$ and $q$ is that they generate the maximal
algebra of operators, invariant under gauge transformations. This algebra,
also referred to as the commutant $(\bw G)'$, is abelian and 12-dimensional:
there are 4 and 3 possible values for $p$ resp.\ $q$.

Global gauge transformations act on the operators $b_i$ as they would act on 
the basis $e_i$:
$$
         \bw g\,b_i\,\bw g^{-1}= \sum_j g_{ji}b_j\qquad (g\in G).
$$
Consequently,
$$
   \bw g\,b_i^*\,\bw g^{-1}=\sum_j(g^*)_{ij}b_j^*\,.
$$ 
The Hodge operator $*$ has already been seen to play a prominent role.
We shall now elaborate on its properties a little further. 
Recall first that $*^2=\one$. Second, we have
$$
   *p*=3-p,\qquad *q*=2-q\,.
$$
Third, the Hodge operator interchanges $b_i$ and $b_i^*$ apart from
possible sign change:
$$
   *b_i* =\kappa b_i^*,\qquad *b_i^**=b_i\kappa \,. 
$$
See (\ref{pari}) for the definition of the parity operator $\kappa$.
As a consequence, we get
$$
     *b_ib_j^**=\kappa b_i^*b_j\kappa= b_i^*b_j =\delta_{ij}-b_jb_i^*\,.
$$
Let $a\in\mbox{End}\,\CC^5$ be arbitrary.
From (\ref{theta}) and the antilinearity of $*$,
\begin{eqnarray*}
     *\theta(a)* &=& \sum_{ij}\bar{a}_{ij}*b_ib_j^**  \\
                 &=& \sum_{ij}(a^*)_{ji}(\delta_{ij}-b_jb^*_i)
                  = \mbox{tr}\,a^*-\theta(a^*)\,.
\end{eqnarray*}
In particular, 
\begin{equation}
  \label{starth}
              *\theta(a)*=\theta(a)\,,\qquad a\in\,{\bf su(5)} 
\end{equation}
owing to the relations $a^*=-a$ and $\mbox{tr}\,a=0$, and ultimately:
\begin{equation}
  \label{cons}
  *\bw g* =\bw g\,,\qquad g=e^a\in G\ .
\end{equation}
This is to demonstrate that we cannot dispense with the trace condition 
$\mbox{tr}\,a=0$, i.e., demanding that $G$ be a subgroup of $SU(5)$
(rather than of $U(5)$)
if we want the consistency condition (\ref{cons}) to be satisfied. 

While elements of the Lie algebra are unchanged under the $*$ operation, their 
Hermitian counterparts $i\theta(a)$ pick up a minus sign owing to 
antilinearity, and so do the charges:
$$   *Q*=-Q,\qquad *Y*=-Y,\qquad *I_3*=-I_3 \ .$$
This is in accord with the conception, formulated before, that the Hodge $*$ 
converts particles into antiparticles and vice versa.

\section{The Higgs Field}

There is no other way than to assume that particles receive their masses 
through the Higgs mechanism. As explained in [2] and [6], the mechanism 
works well only if the Higgs field is an odd operator on the internal 
$\ZZ_2$-graded space. In the present situation, this space is 
$$
             V=\bw\CC^5\otimes\CC^3
$$ 
endowed with the obvious grading $V^\pm=\bw^\pm\CC^5\otimes\CC^3$.
With one Higgs doublet\footnote{Normally, the Higgs field is
thought of as a $(Y=1)$ doublet $(\phi_+,\phi_0)$. We prefer to work instead 
with $\phi_1=\phi_0^*$ and $\phi_2=-\phi_+^*$. Note that $\phi_1$ has zero
electric charge while $\phi_2$ has the charge $Q=-1$. Note also that the scalar
$\phi^*\phi=|\phi_1|^2+|\phi_2|^2$ is gauge invariant.}, 
$$
            \phi=\pmatrix{\phi_1\cr \phi_2\cr}\,,
$$
of hypercharge $Y=-1$, there is more than one choice for the Higgs field 
$\Phi$ (when written in terms of $\phi_1$ and $\phi_2$) if we want $\Phi$ to act as an operator on the (rather large) 
internal space $V$, merely requiring that $\phi$ transforms properly
under the gauge group. As perceived by the founders of the Standard Model,
this freedom of choice shows up in the appearance of a variety of
undetermined Yukawa coupling constants, one for each elementary fermion
field. Traditional thinking forbids to combine the elementary
fermion fields into a single mathematical entity, as the constituents
couple differently to the Higgs doublet $\phi$.

In essence, what we suggest here is to look at the freedom of fixing
parameters (such as Yukawa couplings) from a different perspective. 
Recalling that the commutant 
$(\bw G)'\subset\,\mbox{End}\,\bw\CC^5$ is nontrivial, we are free to choose 
some invariant operator
\begin{equation}
  \label{hin}
               h\in(\bw G)'\otimes\,\mbox{End}\,\CC^3
\end{equation}
and to define the Higgs field by
\begin{equation}
  \label{higgs}
    \Phi(x) =h^*\,\Big(\phi_1(x)b_{4}+\phi_2(x)b_{5}\Big)
\end{equation}
so as to have some field acting on the inner space $\bw\CC^5\otimes\CC^3$.
A gauge transformation, taking $\Phi$ into $\bw g\,\Phi\,\bw g^{-1}$, 
reveals that the complex scalar fields $\phi_i$ transform as desired.

In general, the invariant operator (\ref{hin}) is determined by providing 
12 matrices
\begin{equation}
  \label{matr}
  h(p,q)\in\,\mbox{End}\,\CC^3\qquad  (p=0,\ldots,3,\ q=0,1,2)
\end{equation}
The fact the matrices $h(p,0)$ do not contribute to the Higgs
field (\ref{higgs}) reduces the number 12 to 8. 
A further reduction to 4 is achieved with help of the constraint
\begin{equation}
  \label{star}
                   *\Phi(x)*=\Phi(x)^*       
\end{equation}
implying that particles couple to $\Phi$ in the same way as antiparticles
couple to $\Phi^*$. The condition (\ref{star}) is equivalent to
\begin{equation}
  \label{onec}
                   *h* =h^\dagger
\end{equation}
where another invariant operator $h^\dagger$, associated with $h$, has been 
introduced so as to satisfy the equation $b_ih^\dagger=h^*b_i\kappa$ 
($i=4,5$). 
Written out, the definition is
\begin{equation}
  \label{hpq}
           h^\dagger(p,q)=(-1)^{p+q}h(p,q+1)^*\qquad(p=0,\ldots,3,\,q=0,1).
\end{equation}
Recall that $*h(p,q)*$ means complex conjugation of matrix elements in
conjunction with the replacements $p\to 3-p$ and $q\to 2-q$.
Therefore, another way to write the condition (\ref{onec}) is
\begin{equation}
  \label{prime}
          h(3-p,2-q)=(-1)^{p+q}h(p,q+1)^{T}\qquad(p=0,\ldots,3,\,q=0,1)
\end{equation}
where $A^T$ denotes the transpose of a matrix $A\in\,\mbox{End}\,\CC^3$.
Therefore, among the matrices $h(p,q)$ $(q>0)$, only four have to be fixed  
in order to determine them all. Summarizing:
\begin{quote}\em
We are working with a common Yukawa coupling set equal to unity and
assemble all free parameters, such as the entries of the fermion mass matrix, 
in the operator $h$. 
\end{quote}
Presently, there does not seem to exist a convincing theoretical argument that 
would settle the question as to the origin of $h$ and completely determine 
the matrices $h(p,q)$ entering the mass operator. We need first of all 
understand the mathematical origin of the flavor symmetry (when there is no
Higgs condensate).

Given the form of $h$, the next step is to allege that the Higgs field 
$\Phi$ enters the Dirac operator in a symmetrized form
\begin{equation}
  \label{defL}
       L =i(\Phi+\Phi^*)\in\,\mbox{End}^-V
\end{equation}
so as to satisfy
\begin{equation}
  \label{lstar}
                 L^*=-L\,,\qquad*L*=-L\,.  
\end{equation}
Spontaneous symmetry breaking gives rise to a condensate,
\begin{equation}
              \label{conden}
    \Phi_c= m^* b_{4},\qquad m= r^{1/2} h\,,\qquad
            r=|\langle\phi_1\rangle|^2              
\end{equation}
and hence to a fermion mass operator
\begin{equation}
  \label{mop}
      M =-iL_c= m^*b_{4}+b_4^*m,\qquad*M*=M=M^*,\qquad *m*=m^\dagger\ .
\end{equation}
Note the difference: while the symmetry breaking parameter 
$r$ is but an ordinary constant (to be obtained from the Higgs potential), 
the condensate $\Phi_c$ and the mass operator $M$ are still {\em operators 
of odd type}. In terms of the mass matrices $m(p,q)\in\,\mbox{End}\,\CC^3$
($q>0$) the relation $*m*=m^\dagger$ may be written:
\begin{equation}
  \label{mp}
          m(3-p,2-q)=(-1)^{p+q}m(p,q+1)^{T}\qquad(p=0,\ldots,3,\,q=0,1)
\end{equation}
The spectrum of $M^2$ describes the fermion masses (squared). The components 
$m(p,q)$ of $M$ refer to subgroups of particles, each group
containing three particles of the same electric charge $Q$ but 
different flavors:
$$
\begin{tabular}[t]{llll}
       $m(0,1):$ & $Q=0$   & \qquad $m(0,2):$ & $Q=-1$ \\ 
       $m(2,1):$ & $Q=3/2$ & \qquad $m(2,2):$ & $Q=-1/3$
\end{tabular}
$$
Here, we we have listed only those matrices $m(p,q)$ for which $p$ is even
as the relation (\ref{mp}) says that matrices, for which $p$ is odd, are
related to the former.
At the end, the relation simply guarantees that matter fields ($p=$even)
and antimatter fields ($p=$odd) receive equal masses.

\section{\hspace{-1.3pt}Superconnections, Generalized Dirac Operators,\\ 
         and the Fermionic Action}

Recall the interpretation\footnote{For the present purpose we avoid a 
global bundel-theoretic formulation and shall be content with formulas that
hold locally. This is possible since, locally, all bundels are trivial.}
of the gauge field $A$ as a connection 1-form, i.e., Euclidean spacetime
is modelled by some four-dimensional 
Riemannian manifold $M$, and $A$ takes values in $T^*M\otimes\mbox{Lie}\,G$
where $T^*M$ stands for the cotangent bundle. 
Our main interest lies in the lifted field $\hat{A}=\theta(A)$ taking values in 
$T^*M\otimes\mbox{End}^+\,\bw\CC^5$. In local coordinates,
\begin{eqnarray*}
          A  &=& dx^\mu\otimes A_\mu,\qquad  A_\mu(x)\in\mbox{Lie}\,G\\
      \hat{A}&=& dx^\mu\otimes\hat{A}_\mu,\qquad
                  \hat{A}_\mu(x)\in\mbox{End}^+\,\bw\CC^5\,.
\end{eqnarray*}
A {\em superconnection\/} (see [6] for details) extends the notion of a gauge
connection and is given by some first-order differential operator of odd type,
\begin{equation}
  \label{super}
          \D=D+L\,,\qquad D=d+\hat{A}\,,
\end{equation}
acting on sections\footnote{Sections are referred to as $\bw\CC^5$-valued
differential forms.} of the bundle $\bw T^*M\otimes\bw\CC^5$, where $d$ 
denotes the exterior derivative, $D$ the covariant derivative, and $L$ 
the Higgs field. Generally speaking, $L$ could also include $n$-forms 
($n\ne1$) complying with the oddness\footnote{The property of being odd
or even is defined with reference to the total $\ZZ_2$-grading of 
the space $\bw T^*M\otimes\bw\CC^5$ on which these operators act.}
of $\D$.

With an $2n$-dimensional Riemannian spin${}^c$ manifold [6] one associates 
a Clifford bundle $C(M)$ and a spin bundle which locally coincides
with $M\times S$, $S$ being the {\em spinor space}, a complex
vector space of dimension $2^n$. At $x\in M$, 
there is an isomorphism $c:C(T^*_xM)\otimes\CC \to\,\mbox{End}\,S$ 
and hence a way to construct $\gamma$ matrices,
$$
  \gamma^\mu = c(dx^\mu),\qquad  \{\gamma^\mu,\gamma^\nu\}=-2g^{\mu\nu}\,,
$$
involving $g^{\mu\nu}$, the metric tensor. Spinor fields $\Psi(x)$ take values 
in the space
$$
             F=V\otimes S=\bw\CC^5\otimes\CC^3\otimes S\,.
$$
To use the language of [6], $F$ is a {\em twisted Clifford supermodule} with
$V$ the twisting space.

With a superconnection $\D$ one associates the generalized Dirac operator
$\Dsl$ which, roughly speaking, is obtained from $\D$ by replacing the
basis elements $dx^\mu$ by $\gamma^\mu$ wherever they occur.
Therefore, if $L$ is scalar,
\begin{equation}
  \label{dirac}
          \Dsl =\partial\kern-6pt/ +\hat{A}\kern-6pt/+L =
        \gamma^\mu(\partial_\mu+\hat{A}_\mu)+L\,.
\end{equation}
The Dirac operator acts on spinor fields such that
$$
       \Psi(x)\in F^\pm\qquad\Rightarrow\qquad 
       \Dsl\Psi(x)\in F^\mp\,.
$$
Suppose the manifold $M$ is orientable and $\omega_0$ is a 
volume form. Then the fermionic action, considered as a functional of
the master field $\Psi$, is\footnote{To put a factor $\hf$ in front
is necessary because each elementary fermion field $\psi_I$ enters
the action together with its charge conjugate $\psi^c_I$, both giving
equal contributions.}
\begin{equation}
  \label{sf}
          S_F = \hf\int_M\overline{\Psi}\,i\Dsl\Psi\,\omega_0\,.
\end{equation}
By construction, the integrand incorporates both gauge and Yukawa
interactions. Following [2] we regard $\Psi\mapsto\overline{\Psi}$ as an
antilinear map into the dual space which reverses the chirality:
$\overline{(\Psi_L)}=(\overline{\Psi})_R$.
We also emphasize that, with regard to
functional integration, the (Grassmann) variables $\Psi$ and $\overline{\Psi}$
are dependent, and a reasonable way to write the functional measure is
\begin{equation}
  \label{func}
                d\Psi =\prod_{f,I}d\psi_{fI}
\end{equation}
with $f=1,2,3$ and $I$ running over the subsets of $\{1,2,3,4,5\}$.

Two important features characterize the ansatz (\ref{super}) for the Dirac
operator.
\begin{enumerate}
\item The operator $p$ (but not $q$) commutes with $\Dsl$:
  \begin{equation}
    \label{pcom}
                 p\Dsl =\Dsl p\,.
  \end{equation}
  One consequence is that the baryon number and the lepton number are preserved
  in interactions: leptons and quarks do not couple at vertices of a Feynman
  diagram. Another is that matter is not converted into antimatter.
  Though matter and antimatter may annihilate to yield gauge and Higgs
  particles.
\item The $*$ operator anticommutes with $\Dsl$ and hence commutes with
  $i\Dsl$:
  \begin{equation}
    \label{stard}
                 *(i\Dsl)* =i\Dsl\,.
  \end{equation}
  This fact follows from\footnote{The relation $*\gamma^\mu=-\gamma^\mu*$ is 
  our way of stating that $(\gamma^\mu\psi)^c=-\gamma^\mu\psi^c$ in ordinary
  Dirac theory.} $*\gamma^\mu*=-\gamma^\mu$, (\ref{starth}) and (\ref{lstar}).
  The property (\ref{stard}) of the Dirac operator implies that, if
  $\Psi$ is a generalized Majorana field, so is $\Psi'=i\Dsl\Psi$:
  $$
          \Psi=*\Psi\qquad\Rightarrow\qquad\Psi'=*\Psi'\,,
  $$
  and from
  $$
  \overline{*\Psi}*\!\Psi'=\sum_I\overline{\psi_I^c}\,\psi'_I{}^c
  =\sum_I\overline{\psi'_I}\,{\psi}_I=\overline{\Psi'}\Psi\,,
  $$
  we obtain
  \begin{equation}
    \label{psid}
              \overline{\Psi}\,i\Dsl\Psi=\overline{i\Dsl\Psi}\,\Psi\,.
  \end{equation}
\end{enumerate}
The two properties of the Dirac operator stated above suggest decomposing
$\Psi$ into (anti)matter fields by projecting onto the $p=$even(odd) parts:
\begin{eqnarray*}
             \psi_M+\psi_A&=&\Psi\\
             \psi_M-\psi_A&=&(-1)^p\Psi\,.
\end{eqnarray*}
The fact that the $*$ operator switches between matter and antimatter is
now reflected by the relation $*\psi_M=\psi_A$. The antimatter field is thus
seen to be a redundant variable and, if desired, may be eliminated from the
action functional using
\begin{eqnarray*}
  \overline{\Psi}\,i\Dsl\Psi &=& \overline{\psi_M}\,i\Dsl\psi_M
                                +\overline{\psi_A}\,i\Dsl\psi_A\\
                             &=& \overline{\psi_M}\,i\Dsl\psi_M
                                +\overline{*\psi_M}\,i\Dsl*\psi_M\\
                             &=& \overline{\psi_M}\,i\Dsl\psi_M
                                +\overline{*\psi_M}\,*(i\Dsl\psi_M)
                             \ =\ \overline{\psi_M}\,i\Dsl\psi_M
                                +\overline{i\Dsl\psi_M}\,\psi_M\,.
\end{eqnarray*}
The functional measure (\ref{func}) now assumes the standard form
$d\psi_M d\bar{\psi}_M$.

The matter field may be decomposed even further so as to extract singlet 
and doublet components:
\begin{eqnarray*}
             \psi_{MS}+\psi_{MD}&=&\psi_M\\
             \psi_{MS}-\psi_{AD}&=&(-1)^q\psi_M\,.
\end{eqnarray*}
Owing to the term $L$, part of the Dirac operator and anticommuting with $q$, 
the Dirac operator induces transitions $S\to D$ and $D\to S$. Also, since
$$
              (-1)^{p+q}\psi_M =(-1)^q\psi_M\,,
$$
the field $\psi_{MS}$ is right-handed while $\psi_{MD}$ is left-handed.

\section{Currents}

It is instructive to see how currents emerge from the ansatz (\ref{sf}). 
For this we need only evaluate $\hat{A}\kern-6pt/=\theta(A\kern-6pt/)$ 
in some basis.
Let $(-it_a)_{a=1}^{12}$ be any basis in $\mbox{Lie}\,G$ (so that 
$t_a^*=t_a$). If the gauge field $A$ has real components $A^a_\mu$ given 
by $iA_\mu=A_\mu^at_a$, then 
$$
      i\hat{A}\kern-6pt/=A_\mu^aT_a \gamma^\mu\,,
      \qquad T_a=\theta(t_a)\,.
$$
The representation of gauge couplings in terms of currents is an immediate 
result:
\begin{equation}
  \label{cur}
     \overline{\psi}_M\,i\hat{A}\kern-6pt/\psi_M=j_a^\mu A_\mu^a\,,
     \qquad j_a^\mu= \overline{\psi}_M\,T_a\gamma^\mu\psi_M\,.
\end{equation}
Inspection shows that all currents preserve both $p$ and $q$ in the following sense:
$$
            pT_a =T_ap\,,\qquad qT_a=T_aq
$$
Each current can therefore be decomposed into constituents with a definite
$(p,q)$ assignment,
$$
            j^\mu_a=\sum_{p,q}j^\mu_a(p,q)\,,
$$
meaning that fermions with different $(p,q)$ assignments have no common vertex.
There is no current for the right-handed neutrino: $j^\mu_a(0,0)=0$.

Currents may be characterized according to their behavior
with respect to chirality.
\vspace{5mm}\par\noindent
{\bf Definition}. {\em A current $j_a^\mu=\overline{\psi}_M\,T_a\gamma^\mu
\psi_M$ (for fixed index $a$) is said to be vectorlike if $T_ab_{4}=b_{4}T_a$
and chiral otherwise. Likewise, the interaction $j_a^\mu A_\mu^a$ is said
to be vectorlike resp.\ chiral if the current is.}
\vspace{3mm}

The reason for the above definition is the observation that the operators $b_{4}$
and $b_4^*$ switch between left- and right-handed fields of the same kind.
No matter what gauge group $G$, vectorlike currents are abundant. They form a 
linear subspace of the space of all currents. While the total space is
12-dimensional, the subspace is 9-dimensional for our choice of $G$. It is
spanned by the electric current and eight currents pertaining to the color group.
In other words, the interactions with the photon and the gluons are vectorlike.

There is no unique choice of generators $T_a$ for the 3-dimensional quotient 
space. One choice could be:
\begin{eqnarray}
      I_1 &=&\tfrac{1}{2}(b_{5}b_4^*+b_{4}b_5^*)\nonumber\\
      I_2 &=&\tfrac{i}{2}(b_{5}b_4^*-b_{4}b_5^*)\label{TTT}\\      
      I_3 &=&\tfrac{1}{2}(b_{4}b_4^*-b_{5}b_5^*)\nonumber
\end{eqnarray}
It is indeed easily verified that $I_ab_{4}\ne b_{4}I_a$ ($a=1,2,3$). 
The observation made here corresponds to the fact that
nature provides three basic parity violating interactions of matter
corresponding to the exchange of $W^\pm$ and $Z^0$ bosons. The coupling
to the massive vector bosons specify the selection of $T$'s not commuting 
with $b_{4}$. 
In passing we mention that, with the group $SU(5)$ replacing $G$, the quotient 
space of chiral currents would be 9-dimensional, admitting further parity
violating interactions.

\section{The Bosonic Action}

In the same way as the Yang-Mills connection\footnote{We use the terms
connection and covariant derivative interchangeably.} $D$ gives rise 
to the concept of curvature,
$$
             F= D^2 =\hf dx^\mu\wedge dx^\nu F_{\mu\nu}^a(-iT_a)\,,
$$
the superconnection $\D$ gives rise to a generalized curvature [6],
\begin{equation}
  \label{curv}
      \F =\D^2 =F+DL+L^2\,,
\end{equation}
where we have identified $DL$, the covariant derivative of the Higgs field,
with the supercommutator\footnote{Note that both $D$ and $L$ are odd operators.}
$\sco{D,L}=\{D,L\}$. Without the Higgs field the bosonic action would consist 
of nothing but the Yang-Mills term:
$$
              S_B =\hf\|F\|^2=\hf\int_M |F|^2\omega_0\qquad (L=0).
$$
The precise nature of the invariant $|F|^2$ better be such that it reduces to
$$
   |F|^2= \hf\sum_{a,\mu\nu}(F^a_{\mu\nu})^2
$$
in a flat Euclidean universe. Constructing an invariant $|F|^2$ for 2-forms 
$F$ with the required property is a well-known procedure. 
An extension to $\F$, properly treating $p$-forms of arbitrary order, has been 
given and discussed in [2]. It requires various steps.
\vspace{5mm}\par\noindent
{\bf 1.\ Step}. A basis $-it_a$ in $\mbox{Lie}\,G$ is chosen such that 
\begin{equation}
  \label{scp}
       \mbox{Tr}\,T_aT_b=\delta_{ab}\qquad (a,b=1,\ldots,12)
\end{equation}
where $T_a=\theta(t_a)$ as before.
\vspace{5mm}\par\noindent
{\bf 2.\ Step}. Since any operator $A$ in $\bw\CC^5$ (like $T_a$) extends 
to an operator in
$$
                 V=\bw\CC^5\otimes\CC^3,
$$
we shall not distinguish between $A$ and its extended version, $A\otimes\one$. 
However when it comes to studying their traces, there would
be a distinction unless `Tr' in $\CC^3$ is given another meaning:
$$
  \mbox{Tr}\,C =\oth\sum_{i=1}^3 C_{ii},\qquad C =(C_{ik})\in\mbox{End}\,\CC^3.
$$
The modified trace is but an average. We stipulate that traces,
although written $\mbox{Tr}\,A$, always include an extra factor $\oth$ for 
operators $A$ in $V$. This precaution is necessary in order to preserve
the validity of relations like (\ref{scp}) or $\mbox{Tr}\,q=32$ and many
others.
\vspace{5mm}\par\noindent
{\bf 3.\ Step}. There is no need for a differential form to be homogeneous
(i.e., a $p$-form). As the construction of $\F$ shows, we are dealing here
with very general forms $B$ that are sections of the algebra
$$
           {\cal B}= \bw T^*M\hat{\otimes}\mbox{End}\,V\,.
$$
Since both $\bw T^*M$ and $\mbox{End}\,V$ are superalgebras, so is their
product, and the symbol $\hat{\otimes}$ accounts for that fact: the tensor
product is special for $\ZZ_2$-graded algebras [4]. Any element $B$ taking 
values in ${\cal B}$ has the following structure\footnote{Again, we will be 
content here with a local description.}:
$$
          B=\sum_I dx^I\otimes B_I,\qquad B_I\in\mbox{End}\,V\,.
$$
As $\mbox{dim}\,M=4$,
the sum runs over the subsets $I\subset\{1,2,3,4\}$ and 
$$
        dx^I = dx^{\mu_1}\wedge\cdots\wedge dx^{\mu_r} 
$$
where
$$
   I=\{\mu_1,\ldots,\mu_r\}\,,\qquad \mu_1<\cdots<\mu_r\,,\qquad r=|I|.
$$
We let
\begin{equation}
  \label{bsq}
              |B|^2 =\sum_{I,J} g^{IJ}\,\mbox{Tr}\,B_I^*B_J
\end{equation}
where $g^{IJ}=(dx^I,dx^J)$, to be obtained from the Riemannian metric, and
$g^{IJ}=\delta^{IJ}$ in a flat Euclidean universe.
\vspace{5mm}\par\noindent
Similar to the procedure in [2] we let the gauge-invariant bosonic action 
be given by
\begin{equation}
  \label{bact}
 S_B = \int_M \hf|\F+\mu^2C|^2\omega_0
\end{equation}
with $\mu$ some mass parameter and
\begin{equation}
  \label{cond}
   C\in(\bw G)'\otimes\,\mbox{End}\,\CC^3\,,\qquad C=*C*=C^*.
\end{equation}
Below we shall learn that it suffices to take $C=\one$.
Without the shift operator $\mu^2C$ the minimum of $S_B$ would be attained
for a flat superconnection: $\F=0$. By contrast, the ansatz (\ref{bact}) 
when taken as classical field theory predicts a constant curvature $\F$
in the ground state.

Thanks to the fact that $\F$ splits into $p$-forms 
with $p=0,1,2$, the integrand above consists of three terms only,
$$
  \hf|\F+\mu^2C|^2=\hf|F|^2+ \hf|DL|^2+\hf|L^2+\mu^2C|^2
$$
which are easily identified as the Yang-Mills term, the covariant kinetic
term of the Higgs field, and the Higgs potential.
In order to guarantee the correct behavior of the kinetic term, i.e.,
$$
      \hf|dL|^2=g^{\mu\nu}(\partial_\mu\phi)^*(\partial_\nu\phi),
$$
a normalization condition must be satisfied:
\begin{equation}
  \label{normc}
                  \hf\mbox{Tr}\,qhh^*=1.
\end{equation}
As we shall see (in Section 13), this condition has 
important consequences.

\section{The Higgs Potential and Symmetry Breaking}

In this section, we shall work out the details of the Higgs potential
$$
       V(\phi)=\hf|L^2+\mu^2C|^2=\hf\mbox{Tr}\,(L^2+\mu^2C)^2\,.
$$
Starting from (\ref{higgs}) and (\ref{defL}) we first evaluate some traces
using $*C*=C$, $*\Phi*=\Phi^*$, and the $*$-invariance of the trace:
\begin{eqnarray*}
   \mbox{Tr}\,L^4 &=&2\mbox{Tr}\,(\Phi\Phi^*)^2=
                     (\phi^*\phi)^2\mbox{Tr}\,q(hh^*)^2\\
   \mbox{Tr}\,L^2C&=&-\mbox{Tr}\,(\Phi\Phi^*+\Phi^*\Phi)C=
                     -\mbox{Tr}\,\Phi\Phi^*(C+*C*) \\
               &=&-2\mbox{Tr}\,\Phi\Phi^*C =-\phi^*\phi\,\mbox{Tr}\,qhh^*C\,.
\end{eqnarray*}
As a result of this calculation the Higgs potential may be written in terms
of three constants:
\begin{equation}
  \label{poten}
  V(\phi)=V_0+\frac{\lambda}{4}(\phi^*\phi-r)^2\,.
\end{equation}
Apart from the irrelevant $V_0$, the other two constants are
\begin{equation}
  \label{hcoup}
\begin{tabular}[l]{ll}
  the Higgs coupling  &$\qquad\lambda=2\mbox{Tr}\,q(hh^*)^2$\\ 
  the condensate      &$\qquad r=\mu^2\mbox{Tr}\,qhh^*C/\mbox{Tr}\,q(hh^*)^2$\,.
\end{tabular}
\end{equation}
The invariant operator $C$ is seen to enter the Higgs potential via two 
constants only, $V_0$ and $r$. There will be
no loss of generality when we decide to work $C=\one$ from now on. This
fixes the parameter $\mu$ setting the mass scale, and so 
\begin{equation}
  \label{mu}
                   \lambda r=4\mu^2 
\end{equation}
owing to the normalization condition (\ref{normc}).

The techniques presented here are, for the most part, quite standard.
However, we have decided to work with $r=v^2/2$ where $v$ is, by convention, 
the symmetry breaking parameter used in most texts on the subject.
Note that in the present framework the Higgs coupling constant $\lambda$ 
can never be negative or zero. It cannot be arbitrarily small either. In fact, 
\begin{equation}
  \label{pos}
                    \lambda\ge\frac{1}{4}\ .
\end{equation}
The lower bound can be derived as follows. Define averages
$$
   \langle W\rangle =\mbox{Tr}\,qW/\mbox{Tr}\,q
$$
for operators $W$ acting on $\bw\CC^5\otimes\CC^3$. The obvious inequality
$\langle W^2\rangle\ge\langle W\rangle^2$ for $W=hh^*$ together with
$\mbox{Tr}\,q=32$ and (\ref{normc}) leads to the lower bound for $\lambda$.

As $r>0$, we observe a breakdown of symmetry, and choosing the 
unitary gauge, we have
$$
   \phi_1=2^{-1/2}\varphi+r^{1/2},\qquad\phi_2=0
$$
where $\varphi$ is the neutral scalar Higgs field. To extract its mass, we
need only expand the Higgs potential up to second-order terms:
\begin{equation}
  \label{mH}
          V(\phi)=V_0+\hf m_H^2\varphi^2+O(\varphi^3)\,,\qquad m_H=2\mu\,.
\end{equation}

\section{Masses and Coupling Constants\\ of Vector Bosons}

To give the particulars of vector bosons we must specify
the basis $-it_a$ in $\mbox{Lie}\,G$ obeying (\ref{scp}):
$$
\begin{tabular}[l]{c|c|c|c|}
      & $a=1,2,3$ & $a=4$ & $a=5,\ldots,12$ \\ \hline\vphantom{\Bigg|}
$t_a$ & $\frac{1}{4}\pmatrix{0&0\cr0&\sigma_a\cr}$& 
        $\frac{1}{4}\sqrt{\frac{3}{5}}\pmatrix{\tth\one_3&0\cr0&-\one_2\cr}$&
        $\frac{1}{4}\pmatrix{\lambda_{a-4}&0\cr0&0\cr}$  
\end{tabular}
$$
We have written down the generators $t_a$ as matrices acting on 
$\CC^3\oplus\CC^2$ and made use of the Gell-Mann matrices $\lambda_a$,
which operate on $\CC^3$, and the Pauli matrices $\sigma_a$, which operate
on $\CC^2$. Checking the conditions (\ref{scp}) is facilitated by the
relation [7]
$$  
       \mbox{Tr}\,T_aT_b=8\,\mbox{tr}\,t_at_b\,.
$$
With respect to the given basis, the mass matrix\footnote{A matrix whose
eigenvalues are the masses squared.} $m^2$ of the 12 vector bosons has
matrix elements
\begin{equation}
  \label{vecm}
        m^2_{ab}= \mbox{Tr}\,[T_a,L_c][T_b,L_c]
\end{equation}
(see [2] for a derivation)
with $L_c$, the `condensate of $L$', as in (\ref{mop}). By a straightforward
computation using (\ref{normc}), we obtain
\begin{equation}
  \label{masses}
      m^2_{ab}=r\{t_a,t_b\}_{44}\,,\qquad(a,b=1,\dots,12).  
\end{equation}
where $\{t_a,t_b\}_{ik}$ are the matrix elements of $\{t_a,t_b\}$ as an
operator on $\CC^5$. By inspection, $m^2_{ab}=0$ if either
$a\ge5$ or $b\ge5$ (gluons do not get masses) and we are left with some 
nontrivial $4\times4$ matrix
$$
    (m^2)_{a,b=1}^4=\frac{r}{8}\pmatrix{1&0&0&0\cr0&1&0&0\cr
    0&0&1&-\sqrt{3/5}\cr0&0&-\sqrt{3/5}&3/5\cr}
$$
the eigenvalues of which provide the masses of the four vector bosons
of the electroweak sector:
\begin{equation}
  \label{wzg}
\begin{tabular}[l]{lll}
  mass squared of the $W^\pm$ & : &$m_W^2=r/8$ \\
  mass squared of the $Z^0$     & : &$m_Z^2=r/5$ \\
  mass squared of the $\gamma$& : &0
\end{tabular}
\end{equation}
The eigenvalue $r/8$, when identified with $m_W^2$, together with (\ref{mu})
and (\ref{mH}) leads to an expression for the Higgs coupling constant,
$$
              \lambda =\frac{m_H^2}{8m_W^2}\,,
$$
and together with (\ref{pos}) to a remarkable inequality concerning the 
Higgs mass:
\begin{equation}
  \label{hima}
                  m_H\ge\sqrt{2}\,m_W\,.
\end{equation}
The eigenvectors corresponding to the eigenvalues $r/5$ and 0 are
\begin{eqnarray*}
     t_z&=&\cos\theta_W\, t_3-\sin\theta_W\, t_4,\qquad\cos\theta_W=\sqrt{5/8} \\
     t_0&=&\sin\theta_W\, t_3+\cos\theta_W\, t_4,\qquad\sin\theta_W=\sqrt{3/8}
\end{eqnarray*}
with $\theta_W$ the Weinberg angle. The relation $\sin^2\theta_W=3/8$ is
typical for a $SU(5)$-oriented gauge theory.

The change of basis leads to the photon field $A_\mu^0(x)$ and
the field $Z_\mu(x)$ of the $Z^0$ particle given by 
$Zt_z+A^0t_0=A^3t_3+A^4t_4$.
Hence,
\begin{eqnarray*}
     Z  &=&\cos\theta_W\, A^3-\sin\theta_W\, A^4\\
     A^0&=&\sin\theta_W\, A^3+\cos\theta_W\, A^4
\end{eqnarray*}
Let us now investigate the structure of the basis vector we associate
with the photon:
\begin{eqnarray*}
   t_0 &=&\sqrt{\frac{3}{8}}\ \frac{1}{4}\pmatrix{0&0\cr0&\sigma_3\cr}+
        \sqrt{\frac{5}{8}}\ \frac{1}{4}\sqrt{\frac{3}{5}}
        \pmatrix{\tth\one_3&0\cr0&-\one_2\cr}\\
       &=&\frac{1}{4}\sqrt{\frac{3}{2}}\ \mbox{diag}(\oth,\oth,\oth,0,-1).
\end{eqnarray*}
Physics takes place not in $\CC^5$ but in $\bw\CC^5$.
Letting $T_0=\theta(t_0)$, we find a relationship with the operator of
electric charge,
\begin{equation}
  \label{elch}
         T_0=\frac{1}{4}\sqrt{\frac{3}{2}}\,Q\,.  
\end{equation}
In the same manner we deduce, for $T_z=\theta(t_z)$, 
$$
         T_z=\sqrt{\frac{2}{5}}\left(I_3 -\frac{3}{8}\,Q\right)\,.
$$
Consult (\ref{charge}) and (\ref{TTT}) for the structure of $Q$, $Y$, 
and $I_i$. Together with $2\sqrt{2}T_+=b_{4}b^*_5$ and $2\sqrt{2}T_-=b_{5}b^*_4$
derived from
$$ 
   T_\pm =\frac{1}{\sqrt{2}}(T_1\pm iT_2)=\frac{1}{2\sqrt{2}}(I_1\pm iI_2)
$$
we have thus obtained expressions for all currents of the electroweak sector:
$$
\begin{tabular}[l]{ll}
 $\bar{\psi}_M\gamma^\mu T_0\psi_M\qquad$   & electromagnetic current
                                              (vectorlike)\\
 $\bar{\psi}_M\gamma^\mu T_z\psi_M\qquad$   & neutral weak current (chiral) \\
 $\bar{\psi}_M\gamma^\mu T_\pm\psi_M\qquad$ & charged weak currents (chiral)
\end{tabular}
$$
Consider the fine structure constant $\alpha=e^2/(4\pi)$ and recall the
formula (\ref{elch}). To give the photon the coupling constant $e$ is to say 
that
\begin{equation}
  \label{alph}
    e=\frac{1}{4}\sqrt{\frac{3}{2}}\qquad\mbox{or}\qquad
     \alpha^{-1}=128\pi/3=134.04\ldots
\end{equation}
Renormalization group equations suggest that $\alpha^{-1}$ decreases
slightly with energy. The value above is in accord with this conception:
it lies half way between $\alpha^{-1}(m_Z)=128$ and $\alpha^{-1}(0)=137$.

Sparked by the success, we proceed extracting the coupling constants $g$ 
and $g'$ of the Salam-Weinberg (SW) model from the relation\footnote{Within
the SW formalism, one writes $B$ instead of $A^4$ and calls $\hf Y$ the
hypercharge.}
$$
   \sum_{a=1}^4 A^aT_a=\sum_{i=1}^3 gA^i I_i +\hf g'A^4Y
$$
with the following result:
\begin{equation}
  \label{gg}
  g=\frac{1}{2}\,,\qquad g'=\frac{1}{2}\sqrt{\frac{3}{5}}\,.
\end{equation}
The value $1/2$ obtained for $g$ is consistent with the SW formula
$m_W=gv/2$ taking the relations $r=v^2/2$ and $m_W^2=r/8$ into account.
The value for $g'$ is consistent with the SW formula $g'/g=\tan\theta_W$
provided $\tan\theta_W=\sqrt{3/5}$.

\section{Fermion Masses and CKM Matrices}

Recall the structure of the fermion mass operator $M=M^*$ from Section 7;
$M$ stays invariant under gauge transformations $\bw g$ provided $g$ is
an element of the residual gauge group, leaving $b_{4}$ invariant.
It consists of complex $3\times 3$ mass matrices $m(p,q)$ where $q>0$, 
each one pertaining to a group of three fermions having different flavors
but same quantum numbers other\-wise. The relation $M=*M*$, equivalently the 
relation (\ref{mp}), reflects the matter-antimatter symmetry of $M$. 
A diagonalization can be achieved in the following sense. There is a
unitary operator $U$ on the Hermitian vector space $V\,$, i.e.\ an
element $U\in SU(V)$, such that
\begin{itemize}
\item the transformed mass operator
      $$M'=UMU^*={m'}^*b_{4}+b^*_4m'  $$
      has diagonal mass matrices $m'(p,q)$ for all $p$ and $q\,$,
\item the operator $U$ commutes with gauge transformations from the 
      residual gauge group and may be
      represented as $$  U=U_0b^*_4b_{4}
      +U_1b_{4}b_4^*  $$
      with operators $U_i\in(\bw G)'\otimes\,\mbox{End}\,\CC^3$ where
      $\CC^3$ is the flavor space,
\item the relation $U=*U*$ is satisfied, and hence $U_1
      =*U_0*$,
\item when replacing $T$ by $T'=UTU^*$, we observe that currents such as
      $\bar{\psi}_M\gamma^\mu T\psi_M$ remain unchanged for most generators 
      $T$ except when $T=T_\pm$. The unitary transformation thus
      creates a flavor changing charged current associated with the
      transformed generator
      $$    T'_+=UT_+U^*=T_+\CKM\,,\qquad \CKM=U_1U^*_0$$
      where $\CKM$ is known as the CKM matrix.      
\end{itemize}
We shall now comment on these items. Passage to the diagonal form of all
mass matrices means that we construct the mass eigenstates:
\begin{eqnarray*}
   m'(0,1)&=&\mbox{diag}\,(m_{\nu_e},m_{\nu_\mu},m_{\nu_\tau})\\
   m'(0,2)&=&\mbox{diag}\,(m_{e},m_{\mu},m_{\tau})\\
   m'(2,1)&=&\mbox{diag}\,(m_u,m_c,m_t)\\
   m'(2,2)&=&\mbox{diag}\,(m_d,m_s,m_b)
\end{eqnarray*}
To keep this list short we confined attention to $p=$ even.
The structure of $U$ follows at once from the splitting
$$ V\cong V_0\oplus V_1\,,\qquad V_0\cong V_1$$ 
corresponding to the eigenvalues 0 and 1 of the operator $b_{4}b^*_4$.
This allows us to write
$$
    b_{4}\cong\pmatrix{0&0\cr\one&0\cr}  \,,\quad 
    b_4^*\cong\pmatrix{0&\one\cr0&0\cr}\,,\quad 
    U\cong\pmatrix{U_0&0\cr0&U_1}      \,,\quad 
    M\cong\pmatrix{0&m\cr m^*&0\cr}    
$$
from which all results can be inferred at ease.
The operators $U_0$ and $U_1$ have components 
$$   U_0(p,q)=\cases{\mbox{unitary} & if $q=0,1$\cr 0 & if $q=2$\cr}\qquad
     U_1(p,q)=\cases{\mbox{unitary} & if $q=1,2$\cr 0 & if $q=0$\cr}
$$
and the unitary transformation $M'=UMU^*$ diagonalizes each 
individual mass matrix $m(p,q)$ by means of a 
{\em biunitary\/} transformation: 
$$
          m'(p,q)=U_0(p,q-1)m(p,q)U_1^*(p,q)\qquad (q>0).
$$
Performing these transformations eliminates many irrelevant
parameters from $M$ and hence from the input operator $h$. 

The analysis above also reveals the structure of the CKM matrix:
\begin{enumerate}
\item $\CKM$ is thought of as an element of $(\bw G)'\otimes\,
      \mbox{End}\,\CC^3\,$,
      having components 
      $$   \CKM(p,q)=U_0(p,q)U_1^*(p,q)
           =\cases{\mbox{unitary}& $q=1$\cr 0 & $q=0,2$\cr}   $$
\item As the CKM matrix satisfies the relation $\CKM^*=*U_M*$, 
      we have 
      $$\CKM(3-p,1)=\CKM(p,1)^T\,,\qquad p=0,\ldots,3.$$
\item There are precisely two independent unitary $3\times3$ matrices
      which completely determine $\CKM$:
     \begin{eqnarray*}
       \CKM(0,1) &=&\mbox{CKM matrix of the leptons}\\ 
       \CKM(2,1) &=&\mbox{CKM matrix of the quarks}  
     \end{eqnarray*}
     They enter the charged currents and thus induce flavor changing
     interactions:
      $$  j_+^\mu=\bar{\psi}_M\gamma^\mu T_+\psi'_M\,,\qquad
          j_-^\mu=\bar{\psi}'_M\gamma^\mu T_-\psi_M\,,\qquad
          \psi'_M=\CKM\psi_M\,. $$
     {\em There are no flavor changing neutral currents in such a theory}.
\end{enumerate}
From the fact that traces are invariant under unitary transformations
we obtain the following trace formulas
\begin{eqnarray*}
  \mbox{Tr}\,m(0,1)m(0,1)^* &=&\oth(m^2_{\nu_e}+m^2_{\nu_\mu}+m^2_{\nu_\tau})\\
  \mbox{Tr}\,m(0,2)m(0,2)^* &=&\oth(m^2_{e}+m^2_{\mu}+m^2_{\tau})\\
  \mbox{Tr}\,m(2,1)m(2,1)^* &=&\oth(m^2_u+m^2_c+m^2_t)\\
  \mbox{Tr}\,m(2,2)m(2,2)^* &=&\oth(m^2_d+m^2_s+m^2_b)\,.
\end{eqnarray*}
Consequently, taking (\ref{mp}) into account,
\begin{eqnarray}
  \mbox{Tr}\,M^2&=&\textstyle
            \mbox{Tr}\,qmm^*=\sum_{p,q}q{p\choose3}{q\choose2}
            \mbox{Tr}\,m(p,q)m(p,q)^*\nonumber\\
     &=&\tfrac{4}{3}(m^2_{\nu_e}+m^2_{\nu_\mu}+m^2_{\nu_\tau}
             + m^2_{e}+m^2_{\mu}+m^2_{\tau})\nonumber\\
     & &     +4(m^2_u+m^2_c+m^2_t+m^2_d+m^2_s+m^2_b). \label{sm2}
\end{eqnarray}
In the same manner, one finds
\begin{eqnarray}
  \mbox{Tr}\,M^4&=&\textstyle
            \mbox{Tr}\,q(mm^*)^2=\sum_{p,q}q{p\choose3}{q\choose2}
            \mbox{Tr}\,\bigl(m(p,q)m(p,q)^*\bigr)^2\nonumber\\
     &=&\tfrac{4}{3}(m^4_{\nu_e}+m^4_{\nu_\mu}+m^4_{\nu_\tau}
             + m^4_{e}+m^4_{\mu}+m^4_{\tau})\nonumber\\
     & &     +4(m^4_u+m^4_c+m^4_t+m^4_d+m^4_s+m^4_b). \label{sm4}
\end{eqnarray}
Two equations relate the $W$ mass and the Higgs mass to the fermion masses:
\begin{eqnarray}
             16m_W^2 &=&\mbox{Tr}\,M^2    \label{rel1}\\
          4m_H^2m_W^2 &=&\mbox{Tr}\,M^4\,.\label{rel2}
\end{eqnarray}
The first equality uses (\ref{sm2}), (\ref{conden}), (\ref{normc}), and 
(\ref{wzg}):
$$
  \mbox{Tr}\,M^2=\mbox{Tr}\,qmm^*=r\,\mbox{Tr}\,qhh^*=2r=16m_W^2\,,
$$
while the second uses (\ref{sm4}), (\ref{conden}), (\ref{hcoup}), (\ref{wzg}),
(\ref{mu}), and (\ref{mH}): 
$$
  \mbox{Tr}\,M^4=\mbox{Tr}\,q(mm^*)^2=r^2\mbox{Tr}\,q(hh^*)^2=\hf r^2\lambda 
  =4\lambda rm_W^2=4m_H^2m_W^2\,.
$$
Empirically [8], the mass of the top quark is dominant among the fermion 
masses, and so
$$
      \mbox{Tr}\,M^2\approx 4m_t^2\,,\qquad \mbox{Tr}\,M^4\approx 4m_t^4\,.   
$$
The relations (\ref{rel1}) and (\ref{rel2}) therefore tell us that
$$
          m_H\approx m_t\approx 2m_W\,.
$$
Prior to its observation,
the value of the top quark mass has been predicted [9] on theoretical
grounds within a 10\% error bracket, certainly one of the greatest
triumphs of the Standard Model. Again, the value $2m_W$ for the top mass
obtained above
is off the empirical value $(174\pm7)\,$GeV by 10\%. A value for the Higgs
mass around $2m_W$ has already been predicted in [2]. The fact that $m_H$
depends strongly on the top quark mass, suggesting $m_H\approx 158\,$GeV,
has also been noted by Okumura [10] who argues on the basis of
noncommutative geometry. Similarly, Pirogov and Zenin [11], using
the renormalization group approach, find that a
cutoff equal to the Planck scale would give the Higgs boson a mass
around $160\,$GeV. Surprisingly, the Higgs mass values offered in
the literature, though dependent on very different schemes, all seem to 
to converge.

\vspace{15mm}\par\noindent
{\Large\bf References}
\vspace{5mm}
\begin{enumerate}
\item S.\ Catani et.al.: The QCD and the Standard Model Working Group:\newline
      Summary Report. {\tt hep-ph/0005114}\newline
      M.W.\ Grunewald, Phys.Rept.\ {\bf 322} (1999), 125\newline
      W.\ Hollik, Acta Phys.Polon.\ {\bf B30} (1999), 1787\newline
      C.\ Dionisi, Nucl.Phys.Proc.Suppl.\ {\bf 38} (1995), 125\newline
      S.J.\ Brodsky: Precision Tests of QCD and the Standard Model,\newline
      {\tt hep-ph/9506322}\newline
      P.H.\ Chankowski: Precision Tests of the MSSM, {\tt hep-th/9505304}
\item G.\ Roepstorff: Superconnections and the Higgs Field,\newline
      {\tt hep-th/9801040} and J.Math.Phys.\ {\bf 40} (1999) 2698
\item G.\ Roepstorff: Superconnections and Matter,
      {\tt hep-th/9801045}
\item G.\ Roepstorff: Superconnections: an Interpretation 
      of the Standard Model, {\tt hep-th/9907221}, Electronic Journal of
      Diff.\ Equ., Conf.\ 04, 2000, pp.165-174
\item G.\ Roepstorff and Ch.\ Vehns: An Introduction to Clifford Supermodules,
      {\tt math-ph/9908029}
\item G.\ Roepstorff and Ch.\ Vehns: Generalized Dirac Operators and
      Superconnections, {\tt math-ph/9911006} 
\item G.\ Roepstorff: A Class of Anomaly-Free Gauge Theories,\newline
      {\tt hep-th/0005079} 
\item CDF and DO Collaboration (G.\ Brooijmans for the collaboration):
      Top Quark Mass Measurements at the Tevatron,
      {\tt hep-ex/0005030}
\item E.\ Laenen, J.\ Smith, and W.\ van Neerven, \newline
      Phys.Lett.\ {\bf B321} (1994), 254\newline
      E.L.\ Berger, H.\ Contopanagos, Phys.Rev.\ {\bf D54} (1996), 3085\newline
      S.\ Catani, M.L.\ Mangano, P.\ Nason, L.\ Trentadue, Phys.Lett.\
      {\bf B378} (1996), 329
\item Y.\ Okumura: An estimation of the Higgs boson mass in the two loop
      approximation in a noncommutative differential geometry,\newline
     {\tt hep-ph/9707350}
\item Yu.F.\ Pirogov and O.V.\ Zenin: Two-loop renormalization group profile
     of the standard model and a new generation, {\tt hep-ph/9808414}
\end{enumerate}
\end{document}